\documentclass[10pt,a4paper,twocolumn,aps,prl,notitlepage,nofootinbib,floatfix]{revtex4-1}
\usepackage[LGR,T1]{fontenc}
\usepackage[latin2]{inputenc}
\usepackage{float}
\usepackage{amsmath}
\usepackage{graphicx}
\usepackage{esint}

\makeatletter

\pdfpageheight\paperheight
\pdfpagewidth\paperwidth

\DeclareRobustCommand{\greektext}{%
  \fontencoding{LGR}\selectfont\def\encodingdefault{LGR}}
\DeclareRobustCommand{\textgreek}[1]{\leavevmode{\greektext #1}}
\ProvideTextCommand{\~}{LGR}[1]{\char126#1}

\usepackage{graphicx}
\usepackage{color}
\usepackage[normalem]{ulem}

\makeatother

\begin{document}

\title{Axial gravitational waves in FLRW cosmology and memory effects}

\author{Wojciech Kulczycki}

\author{Edward Malec}

\affiliation{Instytut Fizyki \ Mariana Smoluchowskiego, Uniwersytet Jagiello\'nski,
\L ojasiewicza 11, 30-348 Krak\'ow, Poland}
\begin{abstract}
We show initial data for gravitational axial waves, that are twice
differentiable but which are not $C^{2}$. They generate wave pulses
that interact with matter in the radiation cosmological era. This
forces the radiation matter to rotate. This rotation is permanent
\textemdash{} it persists after the passage of the gravitational pulse.
The observed inhomogeneities of the cosmic microwave background radiation
put a bound onto discontinuities of superhorizon metric perturbations.
We explicitly show that a class of smooth initial metrics that are
at least $C^{2}$ gives rise to gravitational wave pulses that do
not interact with the background during the radiation epoch.
\end{abstract}
\maketitle

\section{Introduction}

Memory effects are known in electrodynamics of infrared fields; if
a test body moves through a pulse of zeroenergy photons, then its
momentum is conserved, but its trajectory will be shifted \cite{1-Staruszkiewicz}
(see commentaries on this effect and later literature in \cite{2-Herdegen}).
In general relativity, Zeldovitch and Polnarev \cite{3-ZP} have shown,
within linearized gravity, that a test body is permanently displaced
by a special type of passing gravitational waves. The characteristic
feature of these effects is the permanent distortion of the metric
after the passage of the (curvature sandwichtype) wave pulses \cite{4-BG,5-Christodoulou,6-BD,7-Thorne}.
This means that these wave pulses are dominated by zero-energy gravitons
(see \cite{8-ZDGH} for a discussion). Usually these are weak signals;
for the detection of GW150914 \cite{9} the memory strain is 20 times
weaker than the oscillatory strain \cite{10}, but under certain circumstances
the memory strain can be much larger than the oscillatory one \cite{11}.

In this paper we discuss the propagation of explicit gravitational
(perturbing metric sandwich-type) waves within radiative Friedmann-Lemaitre-Robertson-Walker
flat space-times. Such gravitational waves do not leave permanent
traces on the relative distance of two test bodies \textemdash{} it
is the same before and after the passage of the pulse. Their Fourier
transforms are not dominated by zero-energy frequencies and the waves
are not dominated by zeroenergy gravitons. Nevertheless, we show that
these waves leave permanent tracks in the radiation background, if
their initial profiles are somewhat deficient in smoothness \textemdash{}
the metric perturbations are twice differentiable, but their second
derivatives are not continuous. In such cases passing wave pulses
force the radiation fluid to rotate.

Let us remark that there is no necessity to assume high orders of
differentiability of solutions of wave equations, even if these equations
are nonlinear. Gravitational wave profiles do not have to be smooth;
they can be even distribution valued. Let us recall the explicit construction
of \textgreek{d}-like gravitational waves by Penrose \cite{12-Penrose1973}
(see also a discussion by Luk and Rodnianski \cite{13-LR}).

We use the Regge-Wheeler formalism. Regge and Wheeler imposed a gauge
condition that allows for the extraction of the two gauge-independent
linearized modes (\textquotedblleft axial\textquotedblright{} and
\textquotedblleft polar\textquotedblright ) of gravitational waves
\cite{14-R-W}. Their original analysis (corrected later by Zerilli
\cite{15-Zerilli}) dealt with gravitational waves in the Schwarzschild
space-time, but their approach appeared convenient in the context
of cosmology \cite{16-MW,17-MWK,18-praca doktorska,19-BP,20-Svitek,21-KM,22-Viaggiu}.
There exists an independent (and equivalent) line of research, with
the use of gauge-invariant quantities \cite{23-GS,24-GMG,25-LTB}.
Clarkson et al. discussed cosmological perturbations in the wider
context of Lemaitre-Tolman cosmologies \cite{25-LTB}; their equations,
restricted to the Friedmann-Lemaitre-Robertson-Walker metrics deformed
by axial modes, coincide with those of \cite{16-MW,21-KM}.

We assume gravitational units $c=8\pi G=1$.

\section{Equations}

We shall use standard coordinates$(\eta,r,\theta,\phi)$, but with
the conformal time coordinate:
\begin{equation}
ds^{2}=a^{2}(\eta)\left(-d\eta^{2}+dr^{2}+r^{2}d\theta^{2}+r^{2}\sin^{2}\theta d\phi^{2}\right).
\end{equation}

\noindent We restrict our attention to the flat $k=0$ universe.

We shall use below the \textquotedblleft conformal\textquotedblright{}
Hubble constant,

\begin{equation}
H(\eta):=\frac{\partial_{\eta}a}{a},
\end{equation}
which is related to the \textquotedblleft ordinary\textquotedblright{}
Hubble constant ${\underline{H}}$ by ${\underline{H}}=\frac{H}{a}$.

Below we put for notational simplicity $Y=Y(\theta):=Y_{l0}(\theta)$
and $Y'=\partial_{\theta}Y$, where $Y_{lm}$ are the spherical harmonics.

The axially perturbed components of the metric in the Regge-Wheeler
gauge read

\begin{equation}
g_{\mu\nu}=a^{2}(\eta)\eta_{\mu\nu}^{(0)}+h_{\mu\nu},\label{metric}
\end{equation}

\noindent where $\eta_{\mu\nu}^{(0)}$ is the Minkowski metric and
the only nonzero components of $h_{\mu\nu}$ are

\begin{equation}
h_{0\phi}=h_{0}\sin\theta~Y'~~~~h_{r\phi}=h_{1}\sin\theta~Y'.\label{axialm}
\end{equation}

\noindent Here $h_{0}=h_{0}(\eta,r)$, $h_{1}=h_{1}(\eta,r)$. The
material field is described by the stress-energy tensor

\begin{equation}
T_{\mu\nu}=(\rho_{0}+p_{0})u_{\mu}u_{\nu}+p_{0}g_{\mu\nu}-\Lambda g_{\mu\nu}.
\end{equation}

It has been already shown in \cite{21-KM} that the mass density $\rho_{0}$
and the pressure $p_{0}$ cannot be disturbed by the axial modes.
Finally, one should allow for the possibility that matter is not necessarily
comoving with the unperturbed cosmological expansion. It appears that
axial modes of the gravitational waves can affect only two components
of the 4-velocity of matter \cite{21-KM}. We have, up to terms linear
in perturbations, 
\begin{eqnarray}
u_{0} & = & -a(\eta),\nonumber \\
u_{\phi} & = & \sin\theta\cdot u(\eta,r)Y'
\end{eqnarray}

\noindent This ensures that $u_{\mu}u^{\mu}=-1$ up to linear terms.

The (background) isotropic and homogeneous solution of Friedmann equations
satisfies the following relations:

\begin{equation}
\rho_{0}=\frac{3}{a^{2}}H^{2}-\Lambda,
\end{equation}
and

\begin{equation}
p_{0}=\Lambda-\frac{1}{a^{2}}H^{2}-\frac{2}{a^{2}}\frac{dH}{d\eta}.
\end{equation}

From these two equations one arrives at

\begin{equation}
\frac{a^{2}}{2}\left(\frac{1}{3}\rho_{0}-p_{0}+\frac{4}{3}\Lambda\right)=H^{2}+\frac{dH}{d\eta}.\label{wyraz_przez_rho_p_i_Lambda}
\end{equation}

We also need (see Sec. IV) the continuity equation

\begin{equation}
\frac{d\rho_{0}}{d\eta}=-3H(p_{0}+\rho_{0}).\label{con}
\end{equation}

Linearized Einstein equations corresponding to the metric (\ref{metric})
read

\begin{equation}
\partial_{r}h_{1}=\partial_{\eta}{h}_{0}\label{mody_osiowe_1}
\end{equation}

\begin{eqnarray}
 &  & \partial_{r}\partial_{\eta}h_{1}-\partial_{r}^{2}h_{0}-2H\partial_{r}h_{1}+\frac{2}{r}\partial_{\eta}{h}_{1}-\frac{4H}{r}h_{1}+\nonumber \\
 &  & +\frac{l(l+1)}{r^{2}}h_{0}=-2a^{3}\left(\rho_{0}+p_{0}\right)u\label{mody_osiowe_2}
\end{eqnarray}

\begin{eqnarray}
 &  & \partial_{\eta}^{2}{h}_{1}-\partial_{r}\partial_{\eta}{h}_{0}-2H\partial_{\eta}h_{1}+\frac{2}{r}\partial_{\eta}h_{0}-2\frac{dH}{d\eta}h_{1}+\nonumber \\
 &  & \frac{l(l+1)-2}{r^{2}}h_{1}=0.\label{mody_osiowe_3}
\end{eqnarray}

Inserting $(\ref{mody_osiowe_1})$ into $(\ref{mody_osiowe_3})$,
defining a new quantity $Q(\eta,r)$ by

\begin{equation}
h_{1}(\eta,r)=ra(\eta)Q(\eta,r),
\end{equation}

\noindent and using (\ref{wyraz_przez_rho_p_i_Lambda}), one gets
\cite{21-KM} 
\begin{equation}
\partial_{\eta}^{2}{Q}-\partial_{r}^{2}Q+\frac{l(l+1)}{r^{2}}Q-\frac{1}{2}a^{2}\left(\frac{1}{3}\rho_{0}-p_{0}+\frac{4}{3}\Lambda\right)Q=0.\label{mody_osiowe_falowe_na_Q}
\end{equation}

The master equation (\ref{mody_osiowe_falowe_na_Q}) can be solved
independently of the remaining equations \textemdash{} it describes
one of the two independent gravitational modes.

From Eq. (\ref{mody_osiowe_1}) one gets $h_{0}$:

\begin{equation}
h_{0}(\eta,r)=A(r)+\int\limits _{\eta_{0}}^{\eta}h'_{1}(\tau,r)d\tau,\label{mody_osiowe_h0}
\end{equation}
where $\eta_{0}$ characterizes the initial hypersurface. The function
$A(r)$ is arbitrary, but if $h_{0}(\eta,r)$ vanishes at the initial
hypersurface, then $A(r)=0$. We assume that $h_{0}(\eta_{0},r)=0$.

\section{Axial gravitational waves and cosmological rotation}

We shall assume the radiation fluid with the equation of state $p_{0}=\frac{\rho_{0}}{3}$
and $\Lambda=0$. Then the conformal factor $a(\eta)=b\eta$, with
$b$ being a constant of the dimension $(\mathrm{length})^{-1}$,
and the energy density $\rho_{0}\propto\eta^{-4}$. We restrict our
attention to quadrupole ($l=2$) modes. During the radiation epoch,
in the absence of the cosmological constant, the quadrupole master
equation reads 
\begin{equation}
\partial_{\eta}^{2}{Q}-\partial_{r}^{2}Q+\frac{6}{r^{2}}Q=0.\label{fala_spelniajaca_zasade_Huygensa}
\end{equation}

Its general solution has the form (see, for instance, \cite{17-MWK})
\begin{equation}
Q(r,\eta)=r^{2}\,{\partial_{r}}\,{\frac{1}{r}}\,{\partial_{r}}\,\bigl(\frac{g+h}{r}\bigr)\label{2.2}
\end{equation}

where functions $g$ and $h$ depend on the combinations $r-\eta$
or $r+\eta$, respectively.

The outgoing gravitational wave is represented by

\begin{equation}
Q(r,\eta)=r^{2}{\partial_{r}\,{\frac{1}{r}}\,{\partial_{r}}\bigl(\frac{g(r-\eta)}{r}}\bigr)\label{2.2b}
\end{equation}

\begin{figure}[H]
\includegraphics[scale=0.65]{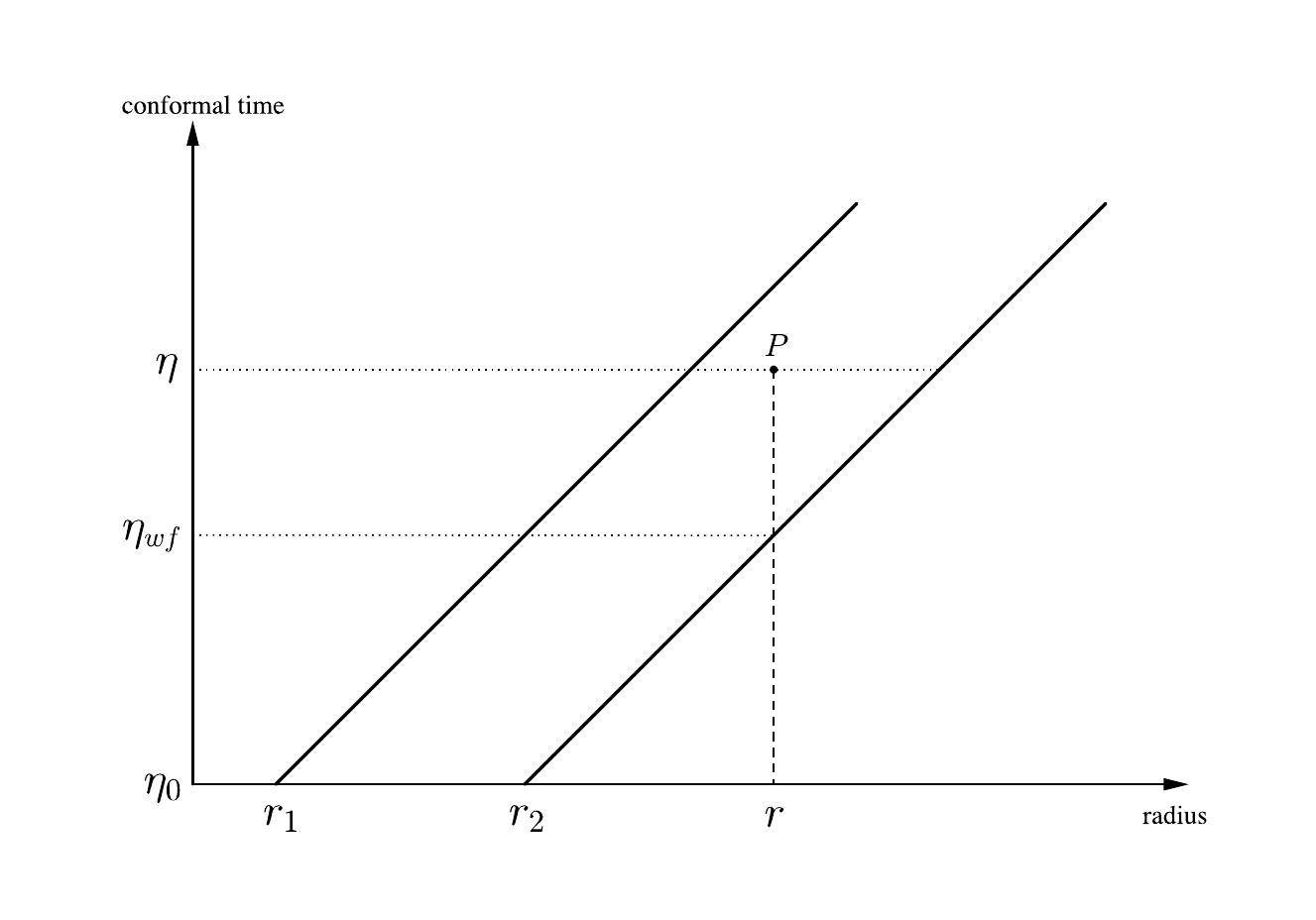}\caption{Space-time diagram showing the propagation of the considered gravitational
pulse ($\eta_{wf}=r-r_{2}+\eta_{0}$). The dashed line shows the integration
contour for the integral in (\ref{mody_osiowe_h0}).}

\label{Fig}
\end{figure}

Assume compact initial data at $\eta=\eta_{0}$, with the initial
support of $g$ contained within the annulus $r_{1}\le r\le r_{2}$.
Let them be at least four times differentiable everywhere except for
isolated points. For the sake of simplicity we assume that there is
only one discontinuity at the wave front. This means that the space-time
metric is at least twice differentiable everywhere with the exception
of the wave front. The wave front of this initial pulse is represented
by the sphere $r=r_{2}+\eta-\eta_{0}$. A straightforward calculation
{[}see Eq. (\ref{mody_osiowe_2}){]} shows that if this wave passes
through a point $P$ of coordinates $(r,\eta)$, with $r>r_{2}$ (see
Fig. \ref{Fig}), then $P$ acquires azimuthal velocity; the quantity

\begin{eqnarray}
u(r,\eta)\rho_{0}a^{3} & = & b\frac{9g(r_{2}-\eta_{0})}{8r^{2}}+b\frac{9\left(r_{2}-\eta_{0}\right)}{8r^{2}}\partial_{\eta}g(r_{2}-\eta)|_{\eta_{0}}-\nonumber \\
 &  & b\frac{9(r-r_{2}+\eta_{0})}{8r}\partial_{\eta}^{2}g(r_{2}-\eta)|_{\eta_{0}}-\nonumber \\
 &  & b\frac{3r}{8}\partial_{\eta}^{3}g(r_{2}-\eta)|_{\eta_{0}}+\nonumber \\
 &  & b\frac{3r}{8}(r-r_{2}+\eta_{0})\partial_{\eta}^{4}g(r_{2}-\eta)|_{\eta_{0}}.\label{u1}
\end{eqnarray}
becomes nonzero. We stress that this effect depends only on characteristics
of the discontinuity of $g$, which is located at the wave front of
the pulse. For smooth initial profiles, with $g\in C^{k}$ $(k\ge4)$,
the right-hand side of (\ref{u1}) vanishes. Notice that the angular
velocity of fluid is proportional to $u$, $\Omega=\frac{u^{\phi}}{u^{0}}=C\frac{u\cos\theta}{ar^{2}}$.
Thus wave profiles, that are at least $C^{4}$, do not enforce cosmological
rotation. Smooth pulses of axial gravitational waves travel through
the cosmological radiative space-time without disturbing the radiation
fluid, in agreement with the assumptions adopted in \cite{16-MW,25-LTB}. 

We work in the linearization regime, which means that perturbed linear
velocities of particles of fluid should be small compared to their
Hubble flow velocities. That implies that we can allow only a mild
discontinuity \textemdash{} that the fourth derivative of $g$ is
finite. Thence the first three derivatives of $g$ must vanish at
the wave front. Then one finds

\begin{eqnarray}
u(r,\eta)\rho_{0}a^{3} & = & b\frac{3r}{8}(r-r_{2}+\eta_{0})\partial_{\eta}^{4}g(r_{2}-\eta)|_{\eta_{0}},\label{u}
\end{eqnarray}

\noindent which is nonzero, if $\partial_{\eta}^{4}g$ has a discontinuity
at the wave front. In such a case the passage of a wave front through
a particle of fluid forces it to rotate around the symmetry axis,
with the angular velocity $\Omega$. This effect is permanent \textemdash{}
the rotation persists after the wave leaves the region.

\section{Discussion}

What kind of physical processes would be responsible for discontinuities
such as those discussed above? In order to gain insight, one would
have to transform the RW perturbations into their standard harmonic
form, with $h_{TT}$ (transversal and traceless) perturbations. In
the first step one defines a new variable $\tilde{\phi}=\phi+X^{\phi}$,
where $X^{\phi}\equiv\frac{Y'\int_{\eta_{0}}^{\eta}d\tau h_{0}}{a^{2}r^{2}\sin\theta}$
(with remaining variables unchanged, $X^{\eta}=X^{r}=X^{\theta}=0$).
The new perturbation metric terms would be found by $\tilde{h}_{\mu\nu}=h_{\mu\nu}-\nabla_{\mu}X_{\nu}-\nabla_{\nu}X_{\mu}$.
The components $\tilde{h}_{0\mu}$ would all vanish, while $\tilde{h}_{r\phi}$
would be shown to fall off as $1/r$.

In the next step one would transform the perturbation metric to the
transversal form$X_{ij}$; its components would be proportional to
$\partial_{\eta}^{2}g$. Assuming that this metric is generated by
weak gravitational wave sources, one would get $X_{ij}\propto\partial_{\eta}^{2}D_{ij}$,
where $D_{ij}$ is the quadrupole moment of the source. Thus we can
conclude that in this case the assumption that $\partial_{\eta}^{2}g_{ij}$
is discontinuous means that the fourth derivatives of the quadrupole
moment are discontinuous. This can happen as a result of the action
of nongravitational forces associated with violent processes. In the
modern cosmological era, the nonsymmetric explosion of a supernova
would constitute an example. If nonsymmetric violent events happen
during the inflation or radiative era, then they would produce gravitational
waves which induce permanent cosmological rotation.

Zones affected by a single pulse of radiation would appear elliptic
on the cosmic microwave background radiation (CMBR) sky. Their size
would depend on the time of formation \textemdash{} pulses created
during the inflation era would be of the order (at least) of the particle
horizon at the recombination era; that is, their angular sizes would
be larger than (roughly) 1 degree. The rotation induced by this pulse
would manifest itself through the Doppler-related perturbations of
the spectra of the CMB radiation, which would be enclosed within the
zone. Its amplitude would be strongest at the boundary of the circular
region and weakest at its center, but the net effect would obviously
depend on the orientation of the symmetry axis of the gravitational
pulse and on the specific behavior of the angular velocity $\Omega$.

In the radiation era the factor $\rho_{0}^{-1}a^{-3}$ linearly increases,
$\rho_{0}a^{3}\propto\eta^{-1}$. The domain of gravitational wave
pulse expands and the radius of its enclosing sphere grows linearly
with the cosmological time, $r\equiv\eta-\eta_{0}+r_{2}$. The linear
rotational velocity of fluid particles is given by $v=ar\sin\theta\Omega=C\frac{u\sin(2\theta)}{2r}$,
where $C=-\frac{3}{2}\sqrt{\frac{5}{\pi}}$. Inserting (\ref{u}),
we get $v\propto\eta(r-r_{2}+\eta_{0})\partial_{\eta}^{4}g(r_{2}-\eta)|_{\eta_{0}}\sin(2\theta)$.
The modulus $|v|$ of linear velocity grows like $\eta$ at a fixed
coordinate location $r$, but it increases like $\eta^{2}$ at points
which are close to the wave front of the travelling pulse of gravitational
radiation. Its amplitude can be estimated by a quantity of the order
of $\frac{|\partial_{\eta}^{4}g(r_{2}-\eta)|_{\eta_{0}}|}{H_{0}^{2}(1+z_{R})^{3}}$,
where $H_{0}$ is the present value of the Hubble escape velocity
and $z_{R}$ is the redshift at the recombination epoch.

The derivation of this bound requires two assumptions. First, we assumed
a simplified evolution picture, where the Universe experiences a sudden
transition from the radiation to the dust dominated epoch at the recombination
time. Second, it is supposed that the gravitational radiation originates
before the end of inflation era.

Combining the above estimate with the known observations of the cosmic
microwave background radiation yields the bound $|\partial_{\eta}^{4}g(r_{2}-\eta)|_{\eta_{0}}|\le10^{-4}{H_{0}^{2}(1+z_{R})^{3}}$;
this is valid for large-scale perturbations.

\section{Summary}

We have found a class of gravitational wave perturbations of radiative
FLRW metrics which induces permanent rotations of the cosmological
background. The smoothness of initial data is the decisive property\textemdash if
initial metrics are at least $C^{2}$ everywhere, then there is no
effect. If the differentiability is lower than $C^{2}$, then passing
gravitational pulses enforce rotation. They might be present in the
primordial (inflationary) gravitational radiation discussed in the
1980s \cite{26} and recently reinvestigated by several researchers
\cite{27}. The observed inhomogeneities of the cosmic microwave background
radiation put a bound onto discontinuities of superhorizon metric
perturbations.

\section{Note added after publication}

This  and the preceding papers (\cite{16-MW} and \cite{21-KM}) show, in particular, that smooth axial gravitational waves  waves do not interact with perfect fluids in FLRW spacetimes. 

The related fact that  axial gravitational waves would not interact with a stationary  
interior of barotropic stars was known at least since 1967. We read in \cite{28a}:  {\it odd-parity motions are not characterized by pulsations which emit gravitational waves; rather, they are characterized by a stationary, differential rotation of the fluid inside the star and by gravitational waves which do not couple to the star at all.}
 In 1991 S. Chandrasekhar and V. Ferrari stated that {\it   the incidence of polar gravitational waves excites fluid motions in the star while the incidence of axial gravitational waves does not}
(\cite{28} and  papers quoted therein).
 
 P.-M Zhang, C. Duval and P. A. Horvathy \cite{29} have shown recently a result similar to \cite{8-ZDGH}, but for impulsive gravitational waves, that after the wave has passed, particles initially at rest move apart with non vanishing constant transverse velocity.

\end{document}